\documentstyle[12pt]{article}

\large
\newcommand{\non}{\nonumber \\}
\newcommand{\be}{\begin{eqnarray}}
\newcommand{\ee}{\end{eqnarray}}

\newcommand{\rar}{\rightarrow}
\newcommand{\ral}{\leftrightarrow}
\newcommand{\rf}[1]{(\ref{#1})}
\newcommand{\erf}[1]{Eq. (\ref{#1})}
\begin{document}
\setlength{\baselineskip}{27pt} 
\pagestyle{empty}
\vfill
\eject
\begin{flushright}
SUNY-NTG-95-44 \\
hep-ph/9511271
\end{flushright}

\vskip 2.0cm
\centerline{\large \bf Master Formula Approach to
$\gamma\gamma \rar \pi\pi$ processes}
\vskip 2.0 cm

\centerline{S. Chernyshev and I. Zahed}
\vskip .4cm
\centerline{\em Department of Physics,
SUNY, Stony Brook, New York 11794-3800}

\vskip 2cm

\centerline{\bf Abstract}
\vskip 3mm
\noindent
We analyze the $\gamma\gamma \rar \pi^0\pi^0$ and
$\gamma\gamma \rar \pi^+\pi^-$ reactions using
the master formula approach to chiral symmetry breaking.
The pertinent vacuum correlators are estimated at tree level, and
the results are compared with one- and two-loop chiral
perturbation theory. The Compton scattering amplitude
and the pion polarizabilities are also discussed.

\vfill
\noindent
\begin{flushleft}
SUNY-NTG-95-44\\
October, 1995
\end{flushleft}
\eject
\pagestyle{plain}
\setcounter{page}{1}

{\bf 1. Introduction}
\bigskip

At low energy, the fusion
process $\gamma\gamma\rar\pi^0\pi^0$ is directly proportional to
pion loops.
One-loop chiral perturbation theory \cite{bij,dhl} yields a
result that is at odds with the data \cite{cball} by several standard
deviations even at threshold, suggesting important correlations in the
scalar-isoscalar channel. Two-loop chiral perturbation \cite{bgs} does
better, with the help of few parameters that are fixed by
resonance saturation. The data can also be fit using constraints from
dispersion theory \cite{pen}, chiral perturbation theory \cite{doho}
and effective models \cite{ko}.

In the present work, we will provide an analysis of
$\gamma\gamma \rar \pi^+ \pi^-$ and $\gamma\gamma\rar \pi^0\pi^0$,
reactions from the point of view of the master formula approach to
chiral symmetry breaking \cite{yz}. In this approach, chiral symmetry
and unitarity are enforced without recourse to an expansion scheme.
The result is a fusion amplitude that is expressed as the sum of
two vacuum correlation functions. These correlation functions are
amenable to power counting, lattice simulations or model calculations.

Using power counting in $1/f_{\pi}$, the various correlation functions may be
analyzed to one-loop. The results have been discussed in \cite{yz},
and are overall similar to one-loop chiral perturbation theory.
The effects of correlations can be either addressed by expanding further the
correlators in $1/f_{\pi}$ or saturating them with physical states. In this
paper, we will choose the latter route, since the former is likely to be
similar to two-loop chiral perturbation theory. In section 2, we present our
calculations. In section 3, we discuss the fusion cross sections and compare
them with one- and two-loop chiral perturbation theory. In section 4, the
Compton amplitudes are derived by crossing and compared to one- and two-loop
chiral perturbation theory. Our concluding remarks are presented in section 5.

\bigskip
{\bf 2. Calculation}
\bigskip

Let us consider the reaction $\gamma^c (q_1) \; \gamma^d (q_2)\rar
\pi^a (k_1) \; \pi^b (k_2)$ in the gauge where the photon polarizabilities
satisfy the condition $\epsilon_{\mu} (q_i ) q^{\mu}_j = 0$, with $i,j=1,2$.
Then, $\epsilon_1\cdot k_1 = - \epsilon_1 \cdot k_2$.
Let us define the Mandelstam variables to be
\be
s &=& (q_1+q_2)^2 = 2q_1\cdot q_2 \non
t &=& (q_1-k_1)^2 = m_{\pi}^2 -2q_1\cdot k_1 \\
u &=& (q_1-k_2)^2 = m_{\pi}^2 -2q_1\cdot k_2 \nonumber
\ee
Throughout $q_1^2=q_2^2=0$ and $p_1^2=p_2^2=m_{\pi}^2$. The master formula
approach to the $\gamma\gamma\rar \pi\pi$ reaction reads \cite{yz}
\be
{\cal T}^{abcd} (s,t,u) =
&+& i \; \epsilon_1\cdot \epsilon_2 \; \bigg( \epsilon^{bce}\epsilon^{eda}
+ \epsilon^{bde} \epsilon^{eca} \bigg) \non
&+& 4i \; \epsilon_1\cdot k_1 \; \epsilon_2\cdot k_2 \; \bigg(
\frac 1{u-m_{\pi}^2} \epsilon^{bcf}\epsilon^{fda} +
\frac 1{t-m_{\pi}^2} \epsilon^{bdf}\epsilon^{fca} \bigg)\nonumber\\
&+&\frac 1{2f_{\pi}^2} \; \epsilon_1^{\mu}\epsilon^{\nu}_2 \;
(k_2-k_1)^{\beta} \;
\epsilon^{abg}\int d^4 y d^4 z e^{-iq_1\cdot y -iq_2\cdot z}
\non &\times&
<0| T^*\bigg( {\bf V}_{\mu}^c (y) {\bf V}_{\nu}^d (z)
{\bf V}_{\beta}^g (0) \bigg) |0>_{conn}  \non
&+&\frac 1{f_{\pi}^2} \; \epsilon_1^{\mu}\epsilon^{\nu}_2
\; k_1^{\alpha} k_2^{\beta}
\int d^4y d^4z_1 d^4 z_2 e^{-iq_1\cdot y +ik_1\cdot z_1 + ik_2\cdot z_2}
\non &\times&
<0| T^*\bigg( {\bf j}_{A\alpha}^a (z_1) {\bf j}_{A\beta}^b (z_2)
{\bf V}_{\mu}^c (y) {\bf V}_{\nu}^d (0) \bigg) |0>_{conn}  \non
&-& \frac {i}{f_{\pi}} \; m_{\pi}^2 \; \delta^{ab} \;
\epsilon_1^{\mu}\epsilon_2^{\nu}
\int d^4 y \; d^4 z \; e^{-iq_1\cdot y -iq_2\cdot z}
\non &\times&
<0| T^*\bigg( {\bf V}_{\mu}^c (y) {\bf V}_{\nu}^d (z) \hat\sigma (0)\bigg)
|0>_{conn}
\label{amp}
\ee
The first and second lines in \rf{amp} are the seagull and Born terms
respectively. ${\bf V}_{\mu}^a$ is the vector current in QCD  and
${\bf j}_{A\alpha}^a$ is the one-pion reduced axial-current in QCD.
The scalar density $\hat\sigma$ is defined by
\be
\hat\sigma = - f_{\pi} -  \frac{\hat m}{f_{\pi}m_{\pi}^2 } \overline{q} q
\ee
For more details on (\ref{amp}) we refer to  \cite{yz}.
The dominant contributions to (\ref{amp}) are shown in Fig. 1.
Note, that for $ c=d=3$ the correlation function ${\bf VVV}$ does not
contribute. For the fusion of two real photons, only the correlation functions
${\bf jjVV}$ and ${\bf VV}\hat\sigma$ contribute. Their contribution to
one-loop was analyzed in \cite{yz}.

In general, the correlation functions
${\bf jjVV}$ and ${\bf VV}\hat\sigma$ diverge at short distances, and require
subtractions. Naive power counting shows that only one-subtraction is needed
for each of the correlation function. The subtraction constants
in ${\bf jjVV}$ and ${\bf VV}\hat\sigma$ will be set
to zero because of charge-conservation in the crossed channel.
Having said this, it is now equivalent to trade the T$^*$-product in
(\ref{amp}) by the T-product, and saturate the time-ordered
correlators with physical states at tree level. Using
\be
<0 | {\bf V}_{\mu}^a (x) |\rho^b (p) > &\sim &
\delta^{ab} \epsilon_{\mu}^V (p) f_{\rho} m_{\rho} e^{-ip\cdot x}
\ee
the result reads
\be
V^{ab}_{\rm tree}  =&+& \epsilon^1 \cdot \epsilon^2 \;
\frac{m^2_\rho f_{\rho}^2}{f_{\pi}^2}
\, \frac i{q_1^2 -m_{\rho}^2} \,
k_1^{\alpha} <\rho^3 (q_1) | \, {\bf j}_{A\alpha}^a \, |A^f (Q)>
\non &\times&
\left ( \frac i{t-m_{A}^2} + \frac i{u-m_{A}^2} \right )\;
k_2^{\beta} <A^f (Q) | \, {\bf j}_{A\beta}^b \, |\rho^3 (q_2) >
\frac i{q_2^2 -m_{\rho}^2} \non
&-& \epsilon^1 \cdot \epsilon^2 \;
\frac {m^2_\rho f_{\rho}^2}{f_{\pi}} m_{\pi}^2 \delta^{ab} \,
\frac i{q_1^2 -m_{\rho}^2} \,
v_{\sigma\rho\rho}(s) \non &\times&
\frac i{s-m_{\sigma}^2} \, <\sigma(q_2)| \hat\sigma |0>
\frac i{q_2^2 -m_{\rho}^2} \label{vab}
\ee
where the propagators carry the Feynman prescription. The state
$|A^f {(Q)}>$ refers to an axial
vector particle of mass $m_A$ with momentum $Q = q_1-k_1$ in the t-channel,
and $Q=q_1-k_2$ in the u-channel, while the state
$|\sigma (q)>$ refers to a scalar particle of  mass $m_\sigma$.
These particles will be assigned specific widths in the discussion to follow.

In Fig. 2, we show additional tree level contributions to ${\bf jjVV}$.
However, these  contributions to the amplitude are two orders of magnitude down
compared to the ones shown in Fig. 1, and will be ignored
\footnote{Note that if we were to express the A-propagator in a Landau-like
gauge, these contributions will just drop from the amplitude because of
transversality.}. Indeed,  the contributions to \rf{amp}
from Fig. 2, are just through the longitudinal parts
$k_1\cdot {\bf j}_A k_2\cdot {\bf j}_A \sim
\big (m_\pi^2/m_A^2 \big )^2$. Thus a factor $m_\pi^2/m_A^2 \sim 10^{-2}$
down compared to the dominant parts in \rf{vab}, following from Fig.1.
We also note that in chiral
models with vector mesons \cite{zb,tensor},
there is usually no $\pi A_1$-mixing,
and so these contributions are just zero in the fusion amplitude.

The matrix elements and vertices appearing in (\ref{vab}) are not known.
Assuming that they
are analytic in the invariant momenta, we will only retain their leading
behaviour at low energy. For the scalar, we will use
 $<0| \overline{q} q |\sigma(q) > = \lambda^2_\sigma \sim m_{\sigma}^2$
as suggested by instanton simulations \cite{shur}, and confirmed by our fit
(see below). For the transition matrix element of the axial-vector current
between the $\rho$ and the $a_1$, we will use the general decomposition
\be
<\rho^3 {(q_1)} | \, {\bf j}_{A\alpha}^a \, |A^f {(q_1-k_1)}> &\sim&
i \epsilon^{3af} \bigg [
k_1^\alpha F_1 \epsilon^A \cdot \epsilon^V +
\epsilon^V_\alpha F_2 k_1 \cdot \epsilon^A +
\epsilon^A_\alpha F_3 k_1 \cdot \epsilon^V  \bigg ]
\label{rja}
\ee
where $\epsilon^A$ and $\epsilon^V$ stand for the vector and
the axial polarizations respectively.
Charge-conservation in the crossed (Compton) channel requires that
$F_1=0$ and $F_2 = -F_3= F_A$. For simplicity, we will set
$F_A$ to  a constant. Finally, for the $\rho\rho\sigma$ vertex, we
choose
\be
v_{\sigma\rho\rho}(s) = i \gamma_{\rho\rho\sigma} \frac{s}{m_\sigma^2}
\label{rrp}
\ee
A constant contribution
to (\ref{rrp}) violates charge neutrality of the $\pi^0$, since
it gives a non-zero charge to $\pi^0$ in the Compton amplitude by crossing.
Order ${\cal O} (s^2)$ have been ignored for simplicity.

Using the above parametrizations for the matrix elements, we obtain for
the fusion amplitude
\be
V^{ab}_{\rm tree}  = i \epsilon^1 \cdot \epsilon^2 \, \delta^{ab} \,
\frac{f_{\rho}^2 m^2_\pi}{f_{\pi}^2 m_\rho^2} \, \bigg \{ \,
F_A^2 \, \frac{s}{m_\pi^2} \frac{g(s,t,u)}{u-m_{A}^2}
+ \frac {\hat m \gamma_{\rho\rho\sigma}}{m^2_\pi}
\frac s{s-m_{\sigma}^2} \,  \bigg \}
\ee
with
\be
g(s,t,u) &=&
\frac{(4m_\pi^2-s)}{2} - \frac {(m_\pi^2+u)^2}{4m_A^2} \non
&&+ \frac {(m_\pi^2-t) (m_\pi^2-u) - s m_\pi^2}{4 m_\rho^2}
+ \frac {s (m_\pi^2+u)^2}{16m_\rho^2 m_A^2}
\label{final}
\ee
The parameters in (\ref{final}) are $F_A$ and
$\gamma_{\rho\rho\sigma}$. The various
masses, decay widths and decay constants $f_A$,$f_{\rho}$, are fixed by
experiment.

\bigskip
{\bf 3. Cross sections}
\bigskip

$\bullet$ For the neutral fusion process
$\gamma\gamma \rar \pi^0\pi^0$ the first (seagull), second (Born)
and third terms in \rf{amp} drop and the amplitude reads
\be
{\cal T}_{\gamma\gamma\rar \pi^0\pi^0} = ie^2 \,
\epsilon_1\cdot\epsilon_2 \,
\frac {f^2_\rho m^2_\pi}{f_{\pi}^2 m^2_\rho}
\frac{ \hat m \gamma_{\rho\rho\sigma}}{m^2_\pi}
\frac s{s-m_{\sigma}^2}
\ee

Let $\Gamma_{\sigma}$ be the momentum dependent width of the scalar particle
\cite{jim},
\be
\Gamma_\sigma (q^2)=\Gamma_1\left({1-4m_\pi^2/q^2\over
1-4m_\pi^2/m_{\sigma}^{\ 2}}\right)^{1/2}
\ee
The differential cross section for the neutral fusion process receives
contribution only from the ${\bf VV}\hat\sigma$ in the form
\be
\bigg(\frac{d\sigma}{d\Omega}\bigg)_{\gamma\gamma\rar \pi^0\pi^0}=
\frac{\alpha^2\beta_V}{4s} \vert \frac{m_{\pi}}{2\alpha} {\bf \alpha}_{\pi}^0
(s) s \vert^2
\label{dnot}
\ee
with a neutral polarization function for the fusion process given by
\be
{\bf \alpha}_{\pi}^0 (s) =
\frac {\alpha m_\pi f^2_\rho}{f_{\pi}^2 m^2_\rho} \,
\frac{ \hat m \gamma_{\rho \rho \sigma}}{ m^2_\pi } \,
\frac 1{ s-m_{\sigma}^2 + i m_{\sigma} \Gamma_{\sigma} }
\label{anot}
\ee
Here $\beta_V = \sqrt{ 1- 4m_{\pi}^2/s}$
is the pion velocity in the CM frame.

Using $m_{\sigma} = 500$MeV and $\Gamma_1 = 550$MeV, an overall fit to the
total cross section as shown in Fig. 2, implies that
\be
\gamma_{\rho \rho \sigma}
\simeq 6.1 \frac {m_{\pi}^2}{\hat m}
\ee
Small changes in the scalar parameters are possible. Large changes, however,
will upset the fit. This implies the presence of a low mass scalar-isoscalar
contribution in the neutral fusion process, albeit with a large width.

$\bullet$
For the charged fusion process
$\gamma\gamma \rar \pi^+\pi^-$ the seagull and Born terms in
\rf{amp} contribute, while the third term ${\bf VVV}$ drops.
The amplitude reads
\be
{\cal T}_{\gamma\gamma\rar \pi^+\pi^-} &=&
- 4ie^2\, \epsilon_1\cdot k_1\, \epsilon_2\cdot k_2 \,
\bigg( \frac 1{t-m_{\pi}^2} + \frac 1{u-m_{\pi}^2} \bigg)\nonumber\\
&& - {2ie^2}\epsilon_1\cdot \epsilon_2
\bigg( 1- \frac{f^2_\rho}{2f_{\pi}^2} \, \frac{s F_A^2}{m_\rho^2}
\, \frac{g(s,t,u)}{u-m_{A}^2}
- \, \frac {f^2_\rho m^2_\pi}{2f_{\pi}^2 m^2_\rho}
\frac {\hat m \gamma_{\rho\rho\sigma} }{m^2_\pi}
\frac s{s-m_{\sigma}^2} \bigg)
\ee
The differential cross section for the charged fusion process
can be written in the following form
\be
\bigg(\frac{d\sigma}{d\Omega}\bigg)_{\gamma\gamma\rar \pi^+\pi^-}=
\frac{\alpha^2\beta_V}{4s}
\bigg(+\vert 1+ \frac{m_{\pi}}{2\alpha} {\bf \alpha}_{\pi}^{\pm }
(s) s \vert^2
+\vert {\bf B} + \frac{m_{\pi}}{2\alpha} {\bf \alpha}_{\pi}^{\pm }
(s) s \vert^2 \bigg)
\label{dplus}
\ee
with a charged polarization function given by
\be
{\bf \alpha}_{\pi}^{\pm} (s) = -
\frac {\alpha m_\pi f^2_\rho }{f_{\pi}^2 m^2_\rho} \,
\bigg(\frac{m_{\pi}^2}s -1\bigg)\,
\bigg ( \, \frac{F_A^2}{m_\pi^2} \, \frac{g(s,t,u)}{u-m_{A}^2} +
\frac {\hat m \gamma_{\rho\rho\sigma} }{m^2_\pi}
\frac 1{s-m_{\sigma}^2} \bigg) \label{aplus}
\ee
and a Born contribution
\be
{\bf B} = -1 + \frac{2sm_{\pi}^2}{{(t-m_{\pi}^2)}{(u-m_{\pi}^2)}}
\label{FIN20}
\ee
For $F_A =1$ the contribution to the Born term is small,
and the differential cross section is in overall agreement with the data as
shown in Fig. 3.

\bigskip
{\bf 4. Compton Scattering}
\bigskip

$\bullet$ For
$\gamma(q_1) \; \pi^0 (k_1) \rar \gamma(q_2) \; \pi^0 (k_2)$
we have by crossing $s\ral t$
\be
\left ( \frac{d\sigma}{d\Omega}
\right )_{\gamma \pi^0 \rar \gamma \pi^0} =
\frac {m^2_\pi}{4s} \; \left | \frac{\alpha m_\pi}{2} \;
\frac{ f^2_{\rho} }{f_{\pi}^2 m^2_{\rho} } \;
\frac{ \hat m \gamma_{\rho \rho \sigma}}{ m^2_\pi } \;
\frac {t}{t-m_{\sigma}^2 + i m_{\sigma} \Gamma_{\sigma} }
\right |^2
\ee
which yields the following neutral pion polarizability
\be
{\bf \alpha}_{\pi}^0 (0) = - \frac{\alpha m_{\pi} }{2} \,
\frac{ f^2_{\rho} } {f_{\pi}^2 m^2_{\rho} } \;
\frac{ \hat m \gamma_{\rho\rho\sigma} }{ m^2_{\pi}} \;
\frac{1}{ m_{\sigma}^2 + \Gamma_{\sigma}^2 } \simeq
- 2.2 \cdot 10^{-4} \, {\rm fm}^3
\ee
where we used
$m_\pi = 135 \ $MeV, $f_\pi = 93 \ $MeV and $f_\rho = 144 \ $MeV.
This result is to be compared with the result of
$-0.49 \cdot 10^{-4}$ fm$^3$ following from one-loop chiral perturbation
theory \cite{polamodel}, and the data
$\alpha_{\pi}^0 ({\rm exp}) = (0.69\pm 0.07\pm 0.04) \cdot 10^{-4}$ fm$^3$
\cite{exp1} and $\alpha_{\pi}^0 ({\rm exp})
= (0.8\pm 2.0) \cdot 10^{-4}$ fm$^3$ \cite{exp2}.
The Compton scattering amplitude as a function of $\sqrt{s}$ is shown in Fig.
4.
(solid line), in comparison with one-loop (dot-dashed) and two-loop (dashed)
chiral perturbation theory \cite{bgs}.

$\bullet$
The differential cross section for the charged Compton process
$\gamma(q_1) \; \pi^\pm (k_1) \rar \gamma(q_2) \; \pi^\pm (k_2)$
is
\be
\bigg(\frac{d\sigma}
{d\Omega}\bigg)_{\gamma\pi^{\pm}\rar \gamma\pi^{\pm}}=
\frac{\alpha^2}{2s}
\bigg(\vert 1+ \frac{m_{\pi}}{2\alpha} {\bf \alpha}_{\pi}^{\pm }
(t) t \vert^2
+\vert \overline{\bf B} + \frac{m_{\pi}}{2\alpha} {\bf \alpha}_{\pi}^{\pm }
(t) t \vert^2 \bigg)
\label{cdplus}
\ee
with a charged polarization function for the charged Compton process given by
\rf{aplus} with $s\rar t$, and a Born contribution
\be
\overline{\bf B} = -1 + \frac{2tm_{\pi}^2}{{(s-m_{\pi}^2)}{(u-m_{\pi}^2)}}
\ee
{}From \rf{cdplus} we read the charged pion polarizability
\be
{\bf \alpha}_{\pi}^{\pm}(0) &=& - \frac{\alpha m_{\pi}}{2} \,
\frac{ f^2_{\rho} }{f_{\pi}^2 m^2_\rho} \, 2 F_A^2 \,
\big [ \; 1- \frac{m_\pi^2}{2m_A^2} \; \big ] \;
\frac{1}{m_\pi^2 -m_{A}^2} - {\bf \alpha}_{\pi}^0 (0) \non
&\simeq& + 2.4 \cdot 10^{-4} \, {\rm fm}^3
\ee
for $F_A = 1$. This result is to be compared with
$\alpha_{\pi}^+ = 2.7 \cdot 10^{-4}$ fm$^3$ following from
chiral perturbation
theory \cite{polamodel}, and the data $\alpha_{\pi}^+({\rm exp})=
(2.2 \pm 1.6) \cdot 10^{-4}$ fm$^3$ \cite{exp1}.

\newpage

\bigskip
{\bf 5. Conclusions}
\bigskip

In the master formula approach to chiral symmetry breaking, the two-photon
fusion process is expressed in terms of two vacuum correlation functions.
One of the correlation function drops in the chiral limit. These correlation
functions are amenable to lattice estimates. In this paper they were saturated
by low mass excitations at tree level.
The results for the charged fusion process are dominated by the Born term,
and overall insensitive to the low mass excitations. In the chargeless
channel, our analysis shows a clear contribution from a broad
scalar-isoscalar resonance with $m_\sigma \sim 500$MeV and
$\Gamma_\sigma \sim 550$MeV, much like the one seen in $\pi\pi$ scattering
for the scalar-isoscalar channel. The pion polarizabilities are
also found in fair agreement with the data.
Our results compare favorably with the two-loop
analysis and results from dispersion theory. In fact, our approximations
are amenable to specific weights in the spectral analysis of the
correlation functions. From this point of view our results are similar
in spirit to the dispersion analysis with full compliance with the
underlying Ward identities. Also they provide simple insights to two
correlation functions that can be compared with future lattice simulations.

\bigskip
{\bf Acknowledgements}
\bigskip

We would like to thank Dr. Hidenaga Yamagishi for discussions.
This work was supported in part by US DOE grant DE-FG-88ER40388.

\newpage
\setlength{\baselineskip}{25pt}

\newpage
\setlength{\baselineskip}{27pt}
\centerline{\Large{\bf Figure captions}}

\bigskip
\bigskip

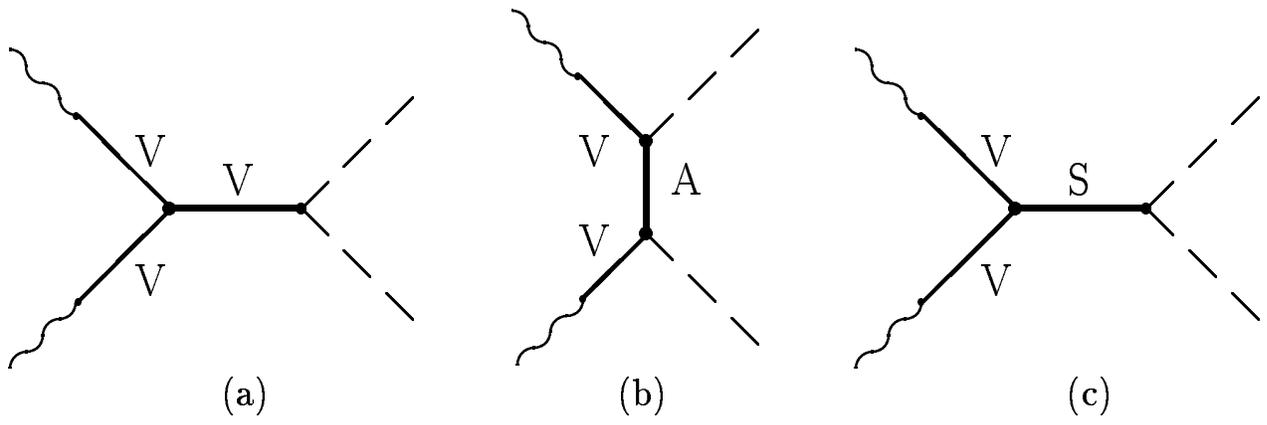
\noindent{\bf Figure 1}:
Dominant contributions to the vacuum correlators in
the process $\gamma\gamma\rar \pi\pi$ as given by
\erf{amp}. The solid V-lines refer to the isovector vector-current ${\bf V}$,
the solid A-lines refers to the one-pion reduced isovector axial-current
${\bf j}_A$, and the solid S-line refers to the scalar current triggered by
$\hat\sigma$. The wiggly lines indicate incoming photons, and the dashed lines
outgoing pions. The crossed contributions are understood.

\bigskip
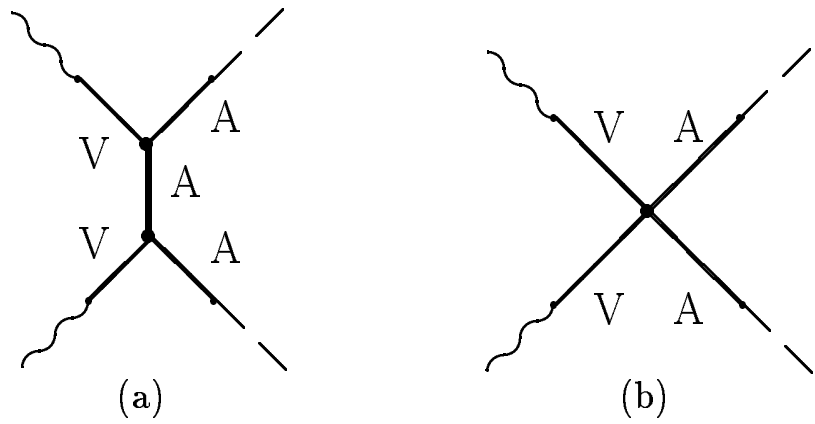
\noindent{\bf Figure 2}:
Subleading diagrams stemming from the vacuum correlator
${\bf jjVV}$ featuring $\pi A_1$-mixing.

\bigskip
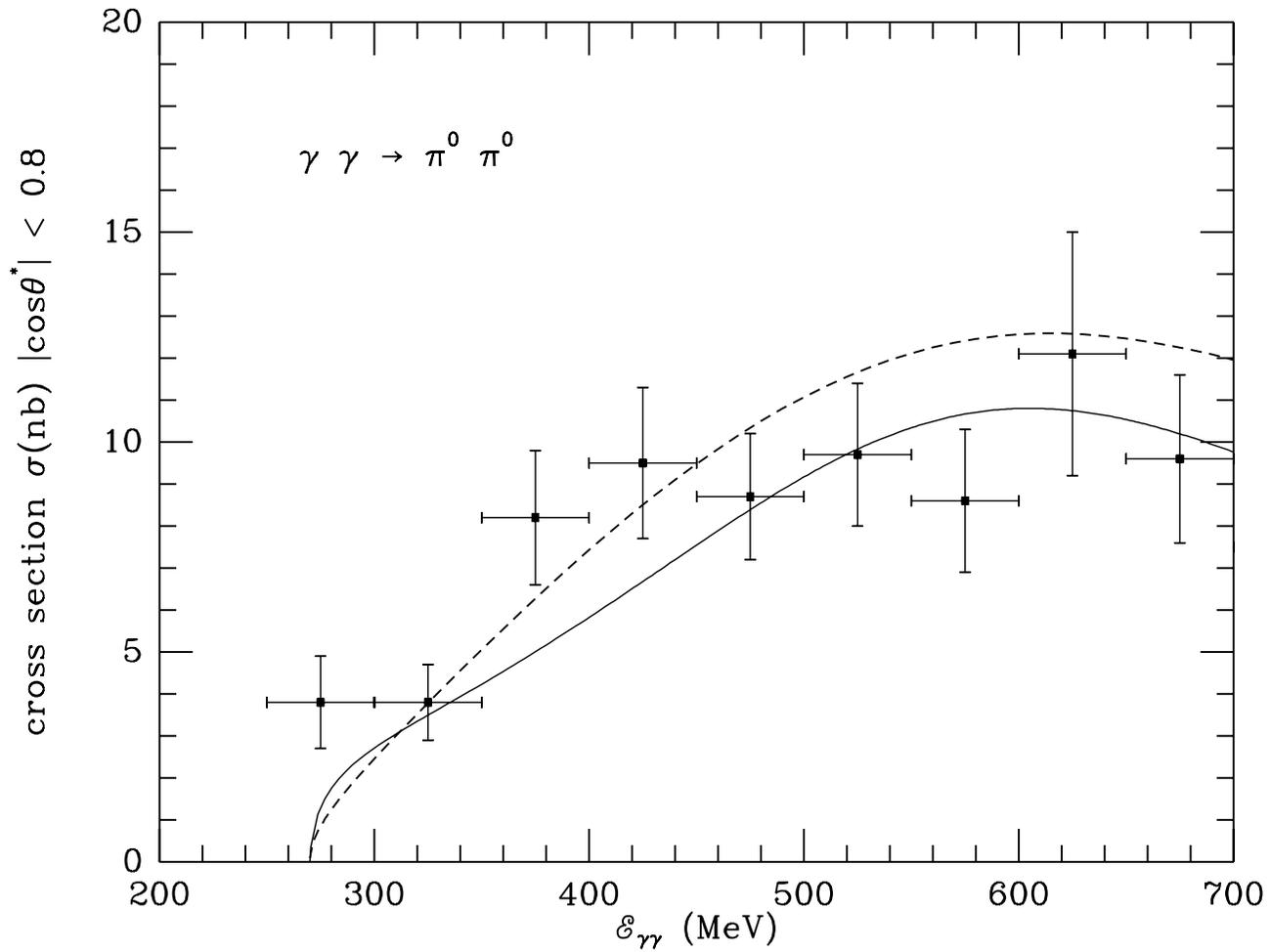
\noindent{\bf Figure 3}:
The $\gamma \gamma \rar \pi^0\pi^0$ cross section
$\sigma$ ($|\cos \theta|\leq 0.8$) as a function of the
center-of-mass energy ${\cal E}_{\gamma \gamma } = \sqrt{s}$ with
the data from the Crystal Ball experiment \cite{cball}.
Our best fit (solid line) as compared to the two-loop result
\cite{bgs} (dashed line).

\bigskip

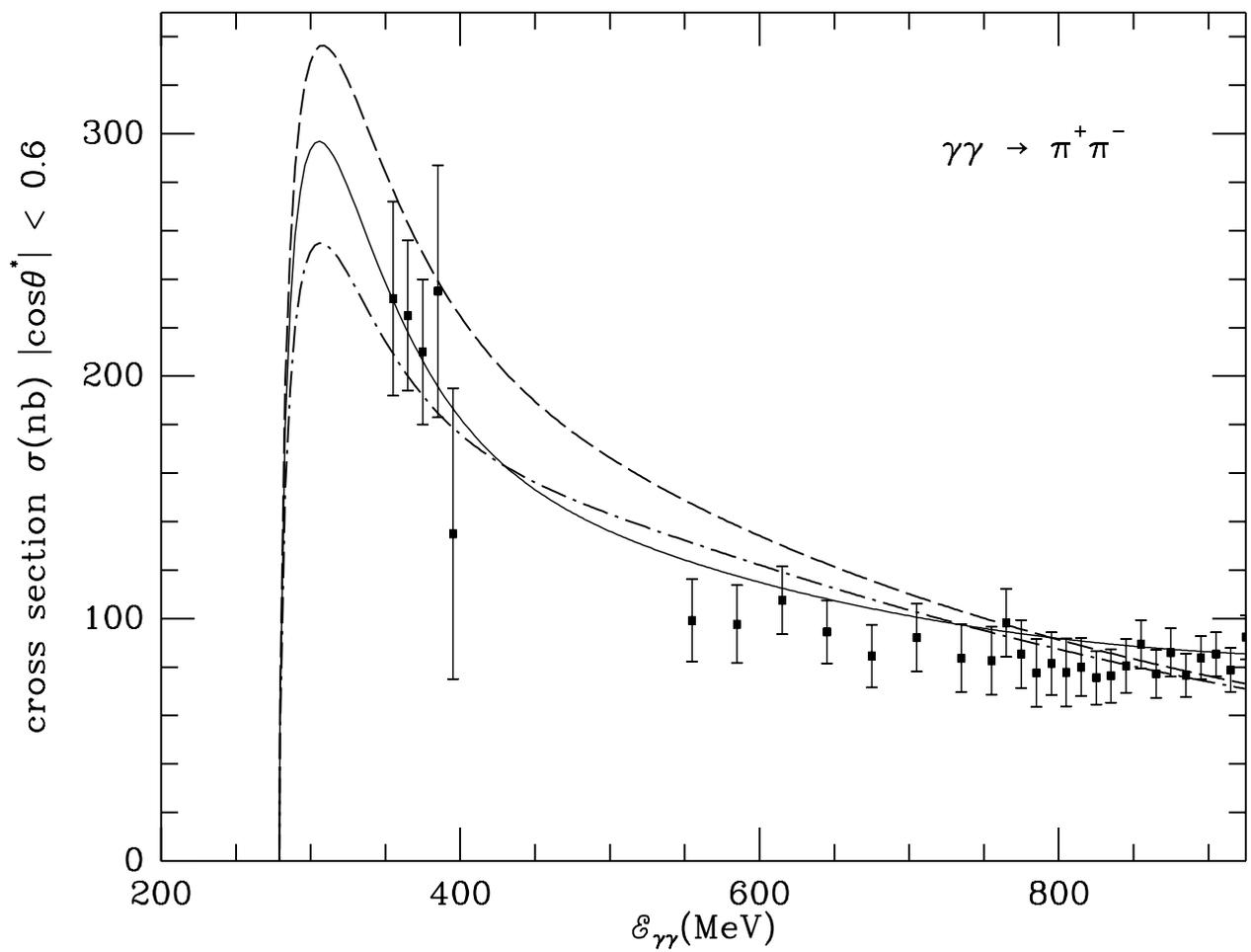
\noindent{\bf Figure 4}:
The $\gamma \gamma \rar \pi^+\pi^-$ process as a function of
${\cal E}_{\gamma \gamma } = \sqrt{s}$.
Our cross section (solid line) versus one-loop $\chi$PT
(dashed) and tree (dot-dashed) \cite{bij,dhl} results.
Data points are taken from Mark II experiment \cite{exp1}.

\bigskip

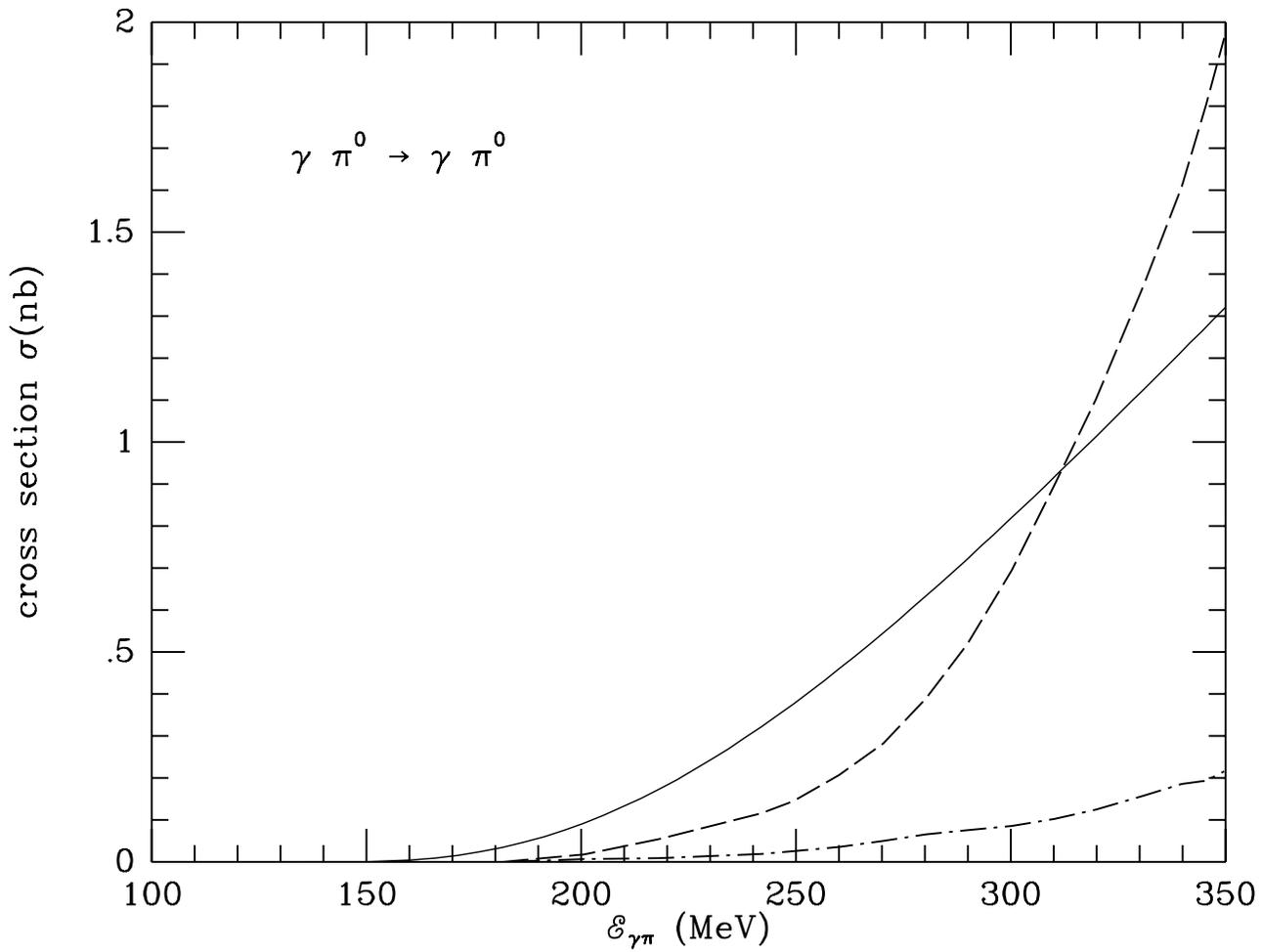
\noindent{\bf Figure 5}:
The Compton scattering $\gamma \pi^\pm \rar \gamma \pi^\pm$
cross section as a function of the CM energy
${\cal E}_{\gamma \pi} = \sqrt{s}$.
Our result (solid line) as compared to the two (dashed) and one-loop
(dot-dashed) $\chi$PT calculations \cite{bgs}.


\begin{thebibliography}{99}

\bibitem{bij}
J. Bijnens and F. Cornet, Nucl. Phys. {\bf B296} (1988) 557.

\bibitem{dhl}
J.F. Donoghue, B.R. Holstein and Y.C. Lin, Phys. Rev. {\bf D37} (1988) 2423.

\bibitem{cball}
The Crystal Ball Collaboration (H. Marsiske et al.),
Phys. Rev. {\bf D41} (1990) 3324.

\bibitem{bgs}
S. Belluci, J. Gasser and M.E. Sainio, Nucl. Phys. {\bf B423} (1994) 80.

\bibitem{pen}
D. Morgand and M.R. Pennington, Phys. Lett. {\bf B272} (1991) 134; \\
T.N. Truong, Phys. Lett. {\bf B313} (1993) 221.

\bibitem{doho}
J.F. Donoghue and B.R. Holstein, Phys. Rev. {\bf D48} (1993) 137.

\bibitem{ko} P. Ko, Phys. Rev. {\bf D41}, 1531 (1990).

\bibitem{yz} H. Yamagishi and I. Zahed, A Master Formula
Approach to Chiral Symmetry Breaking, SUNY-NTG-94-57; Ann. Phys. in Print.

\bibitem{zb} I. Zahed and G.E. Brown, Phys. Rep. {\bf C142} (1986) 1. \\
Ulf--G. Mei{\ss}ner, Phys. Rep. {\bf C161} (1988) 213.

\bibitem{tensor} G. Ecker et al. {\em Nucl. Phys.} {\bf B321} (1989) 311.

\bibitem{shur} E. Shuryak, Acta Phys. Pol. {\bf 25} (1994) 115.

\bibitem{jim} J. Steele, H. Yamagishi, and I. Zahed,
Master Formula Approach to Chiral Symmetry Breaking:
$\pi\pi$-scattering, SUNY-NTG-95-14; Submitted to Nucl. Phys. {\bf B}.

\bibitem{polamodel}
B.R. Holstein, Comm. Nucl. Part. Phys. {\bf 19} (1990) 239,
and references therein.

\bibitem{exp1}
The Mark II Collaboration (J. Boyer et al.),
Phys. Rev. {\bf D42} (1990) 1350. \\
D. Babusci $et$ $al.$, Phys. Lett. {\bf B227} (1992) 158.

\bibitem{exp2}
A.E. Kaloshin and V.V. Serebryakov, Phys. Lett {\bf B278} (1992) 198.

\end{thebibliography}
\end{document}